# Ferromagnetic anomalous Hall effect in Cr-doped $Bi_2Se_3$ thin films via surface-state engineering

*Jisoo Moon[†], Jinwoong Kim[†], Nikesh Koirala[†#], Maryam Salehi[‡], David Vanderbilt[†], Seongshik Oh[†*]*

[†]Department of Physics and Astronomy, Rutgers, The State University of New Jersey, Piscataway, New Jersey 08854, U.S.A.

[‡]Department of Materials Science and Engineering, Rutgers, The State University of New Jersey, Piscataway, New Jersey 08854, U.S.A.

ABSTRACT: The anomalous Hall effect (AHE) is a non-linear Hall effect appearing in magnetic conductors, boosted by internal magnetism beyond what is expected from the ordinary Hall effect. With the recent discovery of the quantized version of the AHE, the quantum anomalous Hall effect (QAHE), in Cr- or V-doped topological insulator (TI) $(Sb,Bi)_2Te_3$ thin films, the AHE in magnetic TIs has been attracting significant interest. However, one of the puzzles in this system has been that while Cr- or V-doped $(Sb,Bi)_2Te_3$ and V-doped $Bi_2Se_3$ exhibit AHE, Cr-doped $Bi_2Se_3$ has failed to exhibit even ferromagnetic AHE, the expected predecessor to the QAHE, though it is the first material predicted to exhibit the QAHE. Here, we have successfully implemented ferromagnetic AHE in Cr-doped $Bi_2Se_3$ thin films by utilizing a surface state engineering scheme. Surprisingly, the observed ferromagnetic AHE in the Cr-doped $Bi_2Se_3$ thin films exhibited only positive slope regardless of the carrier type. We show that this sign problem can be explained by the intrinsic Berry curvature of the system as calculated from a tight-binding model combined with a first-principles method.

The ordinary Hall effect (OHE) is the emergence of a transverse electrical potential difference due to the Lorentz force when an external magnetic field ($H$) is applied perpendicular to the direction of charge flow. The transverse electrical potential difference divided by the longitudinal current is called the Hall resistance ($R_{xy}$). Normally, the $R_{xy}$ grows linearly with the applied magnetic field and the slope is determined by the two-dimensional (2D) sheet carrier density ($n_{2D}$). In magnetic materials, however, the internal magnetism affects the charge motion and the Hall signal can significantly deviate from OHE, sometimes resulting in a hysteretic behavior as a function of $H$; this is called "anomalous" Hall effect (AHE)[1]. Although AHE is commonly observed in ferromagnetic conductors, AHE in magnetic topological insulators (TIs) is special in that it can lead to a quantum anomalous Hall effect (QAHE). Although QAHE was envisioned as early as 1988[2], the first realistic material predicted to exhibit QAHE[3] was a magnetic TI, Cr-doped $Bi_2Se_3$ thin film[4–6]. However, QAHE has been observed only in Cr- and later V-doped $(Bi,Sb)_2Te_3$ thin films[7–12], and it has been challenging to introduce even ferromagnetism (FM) into $Bi_2Se_3$ thin films. The Hall effect measurement in Cr-doped $Bi_2Se_3$ thin films has exhibited only paramagnetic effect without any hysteretic loops[13]. Although V-doped $Bi_2Se_3$ films show hysteretic Hall traces[14], they were orders of magnitude smaller than the quantum value. There have been efforts to reveal the exact origin behind the weakness/absence of FM and AHE in V- and Cr-doped $Bi_2Se_3$ thin films[14–16], but a satisfactory consensus has not been achieved yet.

In order to better understand and specify the problems in the current study, we first compare the Hall traces of the QAHE with positive and negative anomalous Hall conductivity (AHC). Figure 1a shows the characteristic shape of all the observed QAHE so far, in both Cr- and V-doped $(Bi,Sb)_2Te_3$ thin films. In addition to the perfect quantization and the hysteresis, which are the two required features of the QAHE, another common notable feature is that the sign of the AHC is positive, with the slope of the hysteresis loop being positive[7,8]. Even when the hysteresis disappears above the Curie temperature, the zero-field slope still remains positive in all these cases. Moreover, the FM in V-doped $Bi_2Se_3$ exhibited also positive AHC[14].

Figure 1b, on the other hand, shows the shape for a fictitious QAHE with a negative AHC. Such a shape has never been observed in QAHE so far, which has been a mystery.

When a system is close to having a ferromagnetic phase transition, but is actually still on the paramagnetic side of the phase boundary, we may still expect a residual behavior in which the system shows a rapid but continuous change of the magnetization over some small region of $|H|$, with a saturating behavior for larger $|H|$. This behavior should also be reflected in the appearance of a nonlinear Hall conductivity of the kind shown in Fig. 1c and d. In this work, we continue to refer to such a behavior as an "anomalous Hall conductivity" because of the strong nonlinearity, although we emphasize that the system does not display the AHE at zero field.

Figure 1c and d illustrate schematics of variations from an n-type Hall effect curve by positive and negative AHC. The negative slope of the dashed lines indicates the n-type charge carriers in the system. The positive AHC modifies the curve by raising the positive side of the magnetic field and lowering the opposite side, as shown in Fig. 1c. Conversely, the positive side of the magnetic field is pushed down by the negative AHC in parallel with the opposite side being pushed up as shown in Fig. 1d. The same effects appear in p-type curves as in the insets, respectively. Unlike the QAHE in Cr- and V-doped $(Bi,Sb)_2Te_3$ thin films, the Cr-doped $Bi_2Se_3$ thin films are so far reported to exhibit only paramagnetism without a hysteresis loop as in Fig. 1d[13]. If it were to show QAHE, it should be of the negative type as depicted in Fig. 1b. This seemingly important feature has never been discussed before, not to mention that the origin behind this phenomenon remains unknown. This problem and its solution are the focus of the current study presented below.

We first discuss the results for uniformly Cr-doped $Bi_2Se_3$ thin films in Fig. 2. All the films used in the current study are grown on 20 quintuple layers (QLs) of $(Bi_{0.5}In_{0.5})_2Se_3$ (BIS in short) on 20 QLs of $In_2Se_3$ on $Al_2O_3(0001)$ substrates: this buffer layer was previously shown to work as an excellent template for $Bi_2Se_3$ thin films[17–19]. The film structure is sketched in Fig. 2a. Figure 2b shows the Hall effect data for 5%, 7.5% and 10% of uniform Cr doping in 10 QL $Bi_2Se_3$ films. Two distinct effects of the uniform Cr doping in $Bi_2Se_3$ films can be seen in the data. First, it leads to an enhanced $R_{xy}$ with higher slopes at small magnetic fields, resulting in non-linear $R_{xy}(H)$ curves. This indicates that the Cr-doping has an amplification effect on the magnetic field being applied to the $Bi_2Se_3$ films, which is a signature of paramagnetism. Second, it leads to a higher n-type carrier density as manifested in the reduced slope of the room-temperature $R_{xy}(H)$ data in the inset of Fig. 2b. The increased n-type carrier density with Cr doping is likely due to disorders, as can be seen in the degraded reflection high-energy electron diffraction (RHEED) images of heavily-Cr-doped $Bi_2Se_3$ films in Fig. 2e-h. It is well known that almost all defects in $Bi_2Se_3$ act as n-type dopants, thus increasing $n_{2D}$ of the inherent n-type carriers[20].

We hypothesize that there are two factors that give rise to the lack of FM in Cr-doped $Bi_2Se_3$ thin films. The first is the high Fermi level due to disorder from Cr doping and the second is the weakening of the topological character of $Bi_2Se_3$ by the lightness of Cr compared with Bi[13]. Regarding the first, studies of ferromagnetic Cr- or V-doped $(Sb,Bi)_2Te_3$ and V-doped $Bi_2Se_3$ films show that lowering the Fermi level generally enhances the AHE signal[7,14]. Accordingly, it is reasonable to expect that lowering the Fermi level in Cr-doped $Bi_2Se_3$ films should also be helpful for inducing FM as the Fermi level falls into the mass-gap of the surface Dirac cone. Considering that Ca-doping on the BIS buffer layer is known to suppress the n-type carriers[19], we grew two Cr-doped (5% and 10%) $Bi_2Se_3$ films with Ca compensation doping (2%) to see if lowering the Fermi level can induce the desired FM. Figure 2c and d provide the Hall effect data for these films at 5 K. Both films show increased slopes in $R_{xy}(H)$ with the Ca compensation doping, indicating that the n-type $n_{2D}$ is reduced with the Ca doping. However, FM AHE was not observed in either of these two cases. This suggests that reducing only the Fermi level may not be sufficient to induce FM in Cr-doped $Bi_2Se_3$ films.

In order to overcome the second problem, the weakened topological nature of the Bi$_2$Se$_3$ films due to Cr-doping, we adopted the magnetic modulation doping scheme suggested by M. Mogi *et al*[9] who succeeded in increasing the critical temperature for the QAHE through such a scheme. In a similar way, we have sandwiched an 8 QL (Cr-free) Bi$_2$Se$_3$ layer by two heavily (50%) Cr-doped QLs as shown in Fig. 3f. The intent of the surface engineering with the magnetic modulation doping is to maintain the full topological character of the bulk Bi$_2$Se$_3$ film while relying on the outer two heavily-Cr-doped layers for the magnetism. The heavily Cr-doped top and bottom layers are insulating based on resistance measurements on a 50 QL (Cr$_{0.5}$Bi$_{0.5}$)$_2$Se$_3$ film, which is non-measurably insulating even at room temperature. Consequently, we expect that the top and bottom layers provide the Cr-free Bi$_2$Se$_3$ layer with strong magnetism without contributing any conductance on their own. However, as shown in Fig. 3a, this magnetic modulation doping scheme alone does not lead to signatures of FM.

Lastly, we have combined the above two schemes: the Ca compensation doping, to lower the Fermi level, and the magnetic modulation doping, to protect the topological character of the film. This combined method has finally led to FM AHE with a small but clear hysteresis loop as shown in Fig. 3b. Interestingly, when the FM emerges in these surface-state-engineered Cr/Ca-doped Bi$_2$Se$_3$ thin films, the zero-field slope changes from negative to positive. In other words, the FM appears with a positive AHC while the uniformly Cr-doped Bi$_2$Se$_3$ thin films show just magnetic amplification effect without FM signal. This corresponds to a transition from Fig. 1d to Fig. 1c. We observed a similar positive AHC in both 1-8-1 QL and 1-6-1 QL films of Ca$_{0.04}$(Cr$_{0.5}$Bi$_{0.5}$)$_{1.96}$Se$_3$ – Ca$_{0.04}$Bi$_{1.96}$Se$_3$ – Ca$_{0.04}$(Cr$_{0.5}$Bi$_{0.5}$)$_{1.96}$Se$_3$ structure with only the outer most layers doped by 50% Cr for Bi: see Supporting Information Section I for the 1-6-1 QL data.

In Fig. 3c, we tried a higher Ca concentration to lower $n_{2D}$ further. However, the AHE signal vanished possibly due to too much disorder associated with the high Ca concentration[19]. Figure 3d and e show Hall effect data for the surface-engineered films with identical thicknesses (1-8-1 QL) but with different Cr concentration. The film with Cr-25% and Ca-2% shows only p-type Hall effect without any hint of FM, while the other one with Cr-100% and Ca-2% shows a positive AHC, but without a clear hysteresis loop. The lack of the FM for Cr-25% and Ca-2% is probably due to too small content of Cr. The degraded AHE signal for Cr-100% and Ca-2% in Fig. 3e as compared with that for Cr-50% and Ca-2% from Fig. 3b is likely due to poor crystalline feature of the Cr-100% layer as can be seen in the RHEED image of Fig. 2e-h. These results suggest that FM in Cr-doped $Bi_2Se_3$ can develop only within a small window of the Cr and Ca concentrations even in this modulation scheme.

These observations naturally lead to the following three critical questions concerning AHE in Cr-doped $Bi_2Se_3$. (1) Why is it particularly challenging to implement FM in $Bi_2Se_3$ compared with the telluride TIs such as $(Sb,Bi)_2Te_3$? (2) Why does the surface Cr doping help induce FM as compared with the homogeneous Cr doping? (3) Why does a positive AHC emerge when a ferromagnetic order appears? We will address each of three questions below in sequence.

Let's first look into the first question. According to the free-energy argument of the original theoretical study[3], TI can exhibit a ferromagnetic order if $J_{eff}^2 - \chi_L^{-1}\chi_e^{-1} > 0$, where $J_{eff}$ is the effective exchange coupling strength between local magnetic moments and band electrons, and $\chi_L$ and $\chi_e$ are magnetic susceptibilities of the local magnetic moments and the band electrons, respectively. In order to satisfy this condition, we need large $J_{eff}$, large $\chi_L$ and large $\chi_e$. In the original theory, however, the authors focused mostly on $\chi_e$, and showed that magnetic TIs could become ferromagnetic because strong spin-orbit coupling (SOC) would substantially enhance $\chi_e$ through a so-called van Vleck mechanism[21]. According to this picture, $Bi_2Te_3$ is more likely to become ferromagnetic than $Bi_2Se_3$ with the same level of magnetic doping because Te should provide stronger SOC than Se does, and this is consistent with the experimental observation[13]. However, the fact that Cr-doped $Bi_2Se_3$ is not ferromagnetic, contrary to the prediction, implies that the original theory was quantitatively insufficient.

Now the above argument comes to answer the second question. By keeping Cr only in the surface layers, two important conditions are satisfied. First, this keeps the SOC strength of the bulk of $Bi_2Se_3$ almost intact, thus maintaining reasonably high $\chi_e$. Second, it enhances $J_{eff}$ with surface states sitting right next to the magnetic ions providing stronger exchange coupling between the localized, magnetic electrons and the itinerant surface electrons via an RKKY type interaction[22]. There has been debate in the literature as to whether the ferromagnetism observed in magnetic TIs is due to an RKKY[23] vs. a van Vleck mechanism[3,13,24,25], or both[26,27], or neither[15,16]. Our current study suggests that it may not be one or the other, but both, with RKKY enhancing $J_{eff}$ and van Vleck enhancing $\chi_e$ in the $Bi_2Se_3$ thin films.

Before moving on to a theoretical investigation to address the third question above, we first note that the quantized AHC is determined by the quantized Berry phase of occupied bands below an insulating gap, according to

$$\sigma_{AH} = \frac{e^2}{h}(-C).$$

Here $C$ is the Chern number: $C = -1$ leads to a positive AHC as in Fig. 1a, and conversely, $C = +1$ results in a negative AHC, as in Fig. 1b. Although the quantized AHC is easily blurred by either of thermal broadening and/or inaccurate chemical potential, the inherited Hall conductivity should preserve the sign of the AHC before and after a ferromagnetic phase transition. Such is the case on all the previously reported QAHE in Cr- or V-doped $(Bi,Sb)_2Te_3$ and AHE in V-doped $Bi_2Se_3$[7,8]. In this context, even if we are not in the QAH regime, we will still utilize this Berry curvature formalism to determine the sign of the observed anomalous Hall effect.

In order to understand the consistent positive AHC in the Cr- or V-doped $(Bi,Sb)_2(Te,Se)_3$ system, we study a tight-binding model for 8 QLs of $Bi_2Se_3$ with a spatially modulated Zeeman field introduced to represent the leading effect of the Cr doping. Figure 4a and b show the mass gaps (displayed by colors) of 8 QL $Bi_2Se_3$ as a function of the uniform (Fig. 4a) and surface proximity (Fig. 4b) Zeeman fields $\Delta_{Bi}$ and $\Delta_{Se}$ independently applied to the $p$ orbitals of the Bi ($x$-axis) and Se ($y$-axis) to represent the uniform and modulation doping cases. See Supporting Information Section III for more details. For the uniform doping case, a Zeeman field, $H_{ZM} = \frac{1}{2}\Delta_{Bi,Se}\sigma_{\hat{n}}$ is applied along (111) direction, normal to the layers, while for modulation doping, the Zeeman field is modified to take the form $H_{ZM,proximity} = \frac{1}{2}\Delta_{Bi,Se}\sigma_{\hat{n}}\exp\left(-\frac{z}{d}\right)$. Here $\sigma_{\hat{n}}$ is the Pauli matrix where $\hat{n}$ is in the layer stacking (surface normal) direction, $z$ is the depth below the surface, and $d$ = 4 Å is the decay length. A mass-gap of the surface Dirac cone is acquired as a preliminary signature of the transition to the QAH phase induced by the Zeeman fields, and both Figure 4a and b show similar QAH phase boundaries with the right bottom side exhibiting a Chern number of −1. Within these theoretical approximations, then, we find no critical differences between the uniform and modulation doping cases.

We then computed appropriate values of $\Delta_{Bi}$ and $\Delta_{Se}$ by extracting the spin-dependent site energy shifts from a Wannierized model extracted from the first-principles calculations. The results are shown in Fig. 4d, where the Zeeman splittings are significant only inside the Cr-doped QL. Assuming dilute (5.6%) and isotropic Cr doping (Fig. 4c), the values of $\Delta_{Bi}$ = 31 meV and $\Delta_{Se}$ = −6 meV are determined as average Zeeman splittings for each element. Interestingly, the estimated Zeeman splittings for Bi and Se have opposite signs, i.e., positive and negative respectively. However, both atoms act with the same sign in determining the Chern number as shown in Fig. 4a and b. The pair of the calculated Zeeman fields ($\Delta_{Bi}$ = 31 meV, $\Delta_{Se}$ = −6 meV) is marked with a red dot implying the resulting Chern number of −1 in Fig. 4a. The similarity of the mass-gap diagrams in Fig. 4a (uniform) and b (proximity) suggests that the modulation doping case should also have $C$ = −1. Although this theoretical analysis does not explain why we observe FM only with modulation doping, it does show that if we have ferromagnetism with Cr-doped $Bi_2Se_3$, then the resulting sign of the Chern number should be −1, leading to the positive AHC as observed here and previously in other systems[7,8].

In conclusion, we have achieved ferromagnetic AHE in Cr-doped $Bi_2Se_3$ thin films through a combination of magnetic-modulation doping and Fermi level tuning. Surprisingly, the sign of the ferromagnetic AHE was always positive, opposite to that of the paramagnetic and ordinary Hall effect. We show that the sign of the ferromagnetic AHE can be well explained by Berry curvature calculations taking into account the effective Zeeman fields induced by Cr doping. Our study also showcases how the combination of materials engineering and theoretical considerations can lead not only to otherwise inaccessible electronic properties, but also to a deeper understanding of the underlying mechanism.

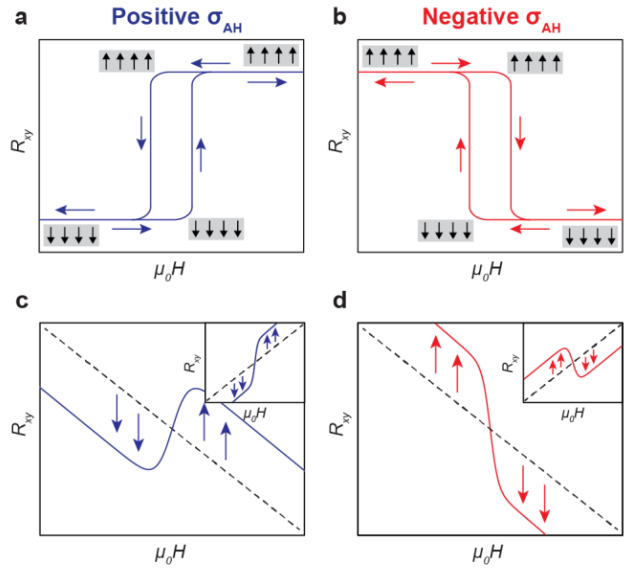

**Figure 1. Schematic comparison of Hall effect curves for magnetically doped TIs.** (a) Characteristic shape of QAHE curves of Cr- and V-doped $(Bi,Sb)_2Te_3$ films with positive AHC. The blue arrows show the direction of the field sweep. (b) The shape for a fictitious QAHE with negative AHC that would be expected for Cr-doped $Bi_2Se_3$ films. Magnetic ordering is shown by cartoons at the corresponding points of the curves, respectively. (c-d) Schematic illustrations of variations of Hall effect curves due to (c) positive and (d) negative AHC. The n-type Hall effect curves are shown by the dashed lines with a negative slope. (c) The positive AHC raises the curve on the positive side of the external magnetic field $H$, and lowers it on the opposite side as indicated by the blue arrows. (d) The negative AHC, on the contrary, lowers the curve on the positive side, and raises it on the other side. The insets show the same variations from a p-type Hall effect curve.

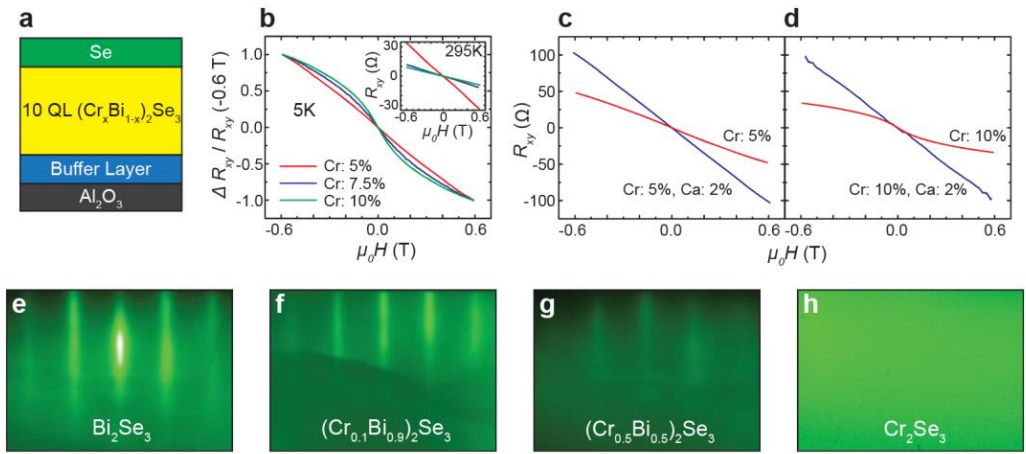

**Figure 2. Hall effect data for uniformly Cr-doped Bi$_2$Se$_3$ films and related RHEED images.** (a) Schematic of the layered structure of uniformly Cr-doped Bi$_2$Se$_3$ films (b) Hall effect data for 5%, 7.5% and 10% uniform Cr doping. $\Delta R_{xy}(H) / R_{xy}(-0.6\ \text{T})$ is plotted instead of $R_{xy}(H)$ for simplicity. The inset shows the Hall effect results at room temperature. (c-d) Comparison of Hall effect data with and without Ca doping with (c) 5% and (d) 10% Cr doping. (e-h) RHEED images of (e) pure Bi$_2$Se$_3$; (f) 10% and (g) 50% Cr-doped Bi$_2$Se$_3$; and (h) Cr$_2$Se$_3$ films. As the Cr concentration increases, the RHEED pattern becomes faint, eventually vanishing at 100% of Cr. The dark areas in (f) and (g) are due to an artifact on the RHEED screen.

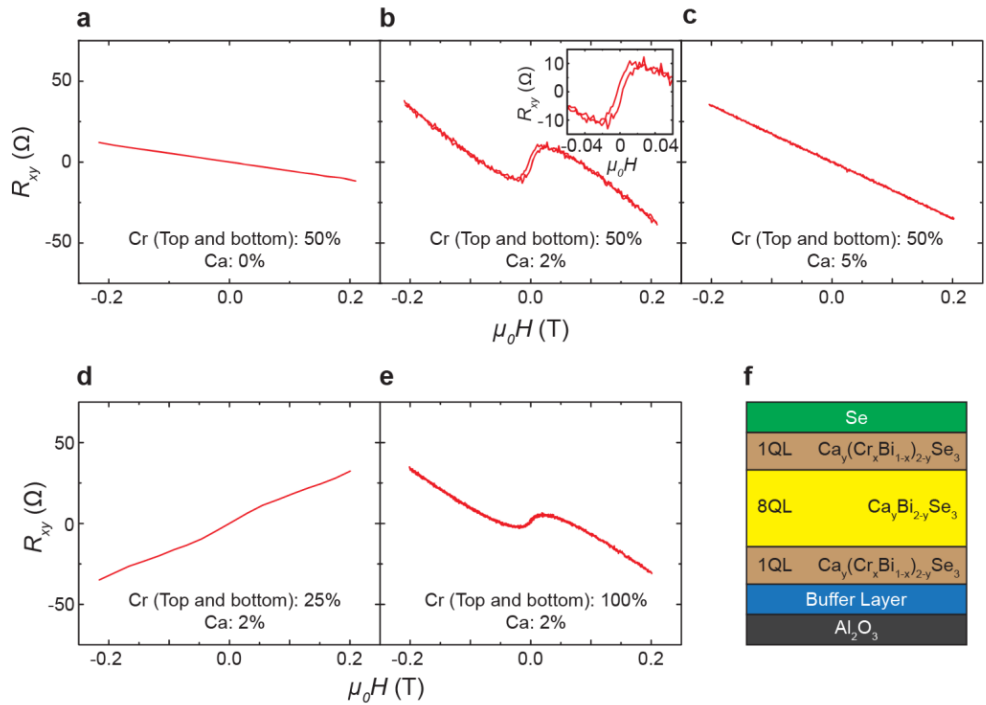

**Figure 3. Hall effect data for surface-state-engineered Cr-doped Bi$_2$Se$_3$ films.** (a-c) Hall effect data in the Bi$_2$Se$_3$ films with 50% of Cr doping only at the top and bottom layer layers, with (a) 0%, (b) 2% and (c) 5% of Ca doping. The inset of (b) presents zoomed-in data near zero magnetic field to show the hysteresis loop clearly. (d-e) Hall effect data in the Bi$_2$Se$_3$ films with (d) 25% and (e) 100% of Cr doping at the top and bottom layers, with 2% of Ca doping. (f) Schematic of the layered structure.

Page 13 of 21

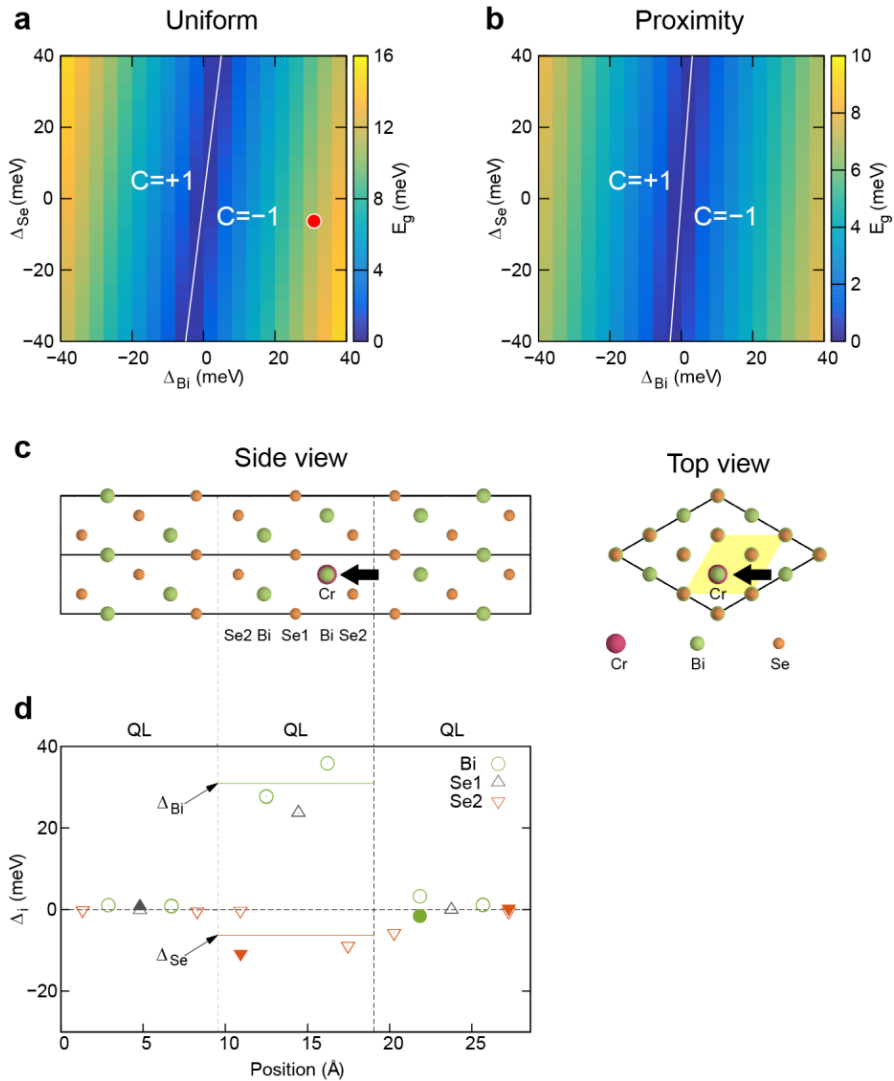

**Figure 4. Model study of the QAH phase of Cr-doped Bi$_2$Se$_3$.** (a-b) Calculated Dirac cone mass gap for 8 QLs of Bi$_2$Se$_3$ under Zeeman field on Bi and Se atoms for (a) uniform field (homogeneous doping), and (b) proximity field (modulation doping). The Zeeman field on the Bi (Se) site induces a Chern number of −1 (+1) for both cases. Red dot shows the position of the averaged Zeeman field in (d). (c) A side and top view of a $\sqrt{3} \times \sqrt{3} \times 1$ hexagonal supercell structure of 3 QLs of Bi$_2$Se$_3$ with one Cr atom substituting for a Bi atom as marked by an arrow. Black solid lines denote the periodic cell. Large, medium, and small spheres indicate Cr, Bi, and Se atoms, respectively. The shaded area on the top view illustrates the in-plane primitive unit cell. (d) Calculated Zeeman splitting of each basis orbital in the supercell structure. Horizontal solid lines are the averaged Zeeman field of the Bi and Se2 atoms, respectively, inside the Cr-doped QL. Closed symbols denote atoms located at the same *xy* coordinates as the Cr atom.

REFERENCES

(1) Nagaosa, N.; Sinova, J.; Onoda, S.; MacDonald, A. H.; Ong, N. P. Anomalous Hall Effect. *Rev. Mod. Phys.* **2010**, *82* (2), 1539–1592.

(2) Haldane, F. D. M. Model for a Quantum Hall Effect without Landau Levels: Condensed-Matter Realization of the "Parity Anomaly." *Phys. Rev. Lett.* **1988**, *61* (18), 2015–2018.

(3) Yu, R.; Zhang, W.; Zhang, H.-J.; Zhang, S.-C.; Dai, X.; Fang, Z. Quantized Anomalous Hall Effect in Magnetic Topological Insulators. *Science* **2010**, *329* (5987), 61–64.

(4) Zhang, H.; Liu, C.-X.; Qi, X.-L.; Dai, X.; Fang, Z.; Zhang, S.-C. Topological Insulators in $Bi_2Se_3$ $Bi_2Te_3$ and $Sb_2Te_3$ with a Single Dirac Cone on the Surface. *Nat. Phys.* **2009**, *5* (6), 438–442.

(5) Hor, Y. S.; Richardella, A.; Roushan, P.; Xia, Y.; Checkelsky, J. G.; Yazdani, A.; Hasan, M. Z.; Ong, N. P.; Cava, R. J. P-Type $Bi_2Se_3$ for Topological Insulator and Low-Temperature Thermoelectric Applications. *Phys. Rev. B* **2009**, *79* (19), 195208.

(6) Hsieh, D.; Xia, Y.; Wray, L.; Qian, D.; Pal, A.; Dil, J. H.; Osterwalder, J.; Meier, F.; Bihlmayer, G.; Kane, C. L.; Hor, Y. S.; Cava, R. J.; Hasan, M. Z. Observation of Unconventional Quantum Spin Textures in Topological Insulators. *Science* **2009**, *323* (5916), 919–922.

(7) Chang, C. Z.; Zhang, J. S.; Feng, X.; Shen, J.; Zhang, Z. C.; Guo, M. H.; Li, K.; Ou, Y. B.; Wei, P.; Wang, L. L.; Ji, Z. Q.; Feng, Y.; Ji, S. H.; Chen, X.; Jia, J. F.; Dai, X.; Fang, Z.; Zhang, S. C.; He, K.; Wang, Y. Y.; Lu, L.; Ma, X. C.; Xue, Q. K. Experimental Observation of the Quantum Anomalous Hall Effect in a Magnetic Topological Insulator. *Science* **2013**, *340* (6129), 167–170.

(8) Chang, C.-Z.; Zhao, W.; Kim, D. Y.; Zhang, H.; Assaf, B. A.; Heiman, D.; Zhang, S.-C.; Liu, C.; Chan, M. H. W.; Moodera, J. S. High-Precision Realization of Robust Quantum Anomalous Hall

State in a Hard Ferromagnetic Topological Insulator. *Nat. Mater.* **2015**, *14* (5), 473–477.

(9) Mogi, M.; Yoshimi, R.; Tsukazaki, A.; Yasuda, K.; Kozuka, Y.; Takahashi, K. S.; Kawasaki, M.; Tokura, Y. Magnetic Modulation Doping in Topological Insulators toward Higher-Temperature Quantum Anomalous Hall Effect. *Appl. Phys. Lett.* **2015**, *107* (18), 182401.

(10) Feng, X.; Feng, Y.; Wang, J.; Ou, Y.; Hao, Z.; Liu, C.; Zhang, Z.; Zhang, L.; Lin, C.; Liao, J.; Li, Y.; Wang, L. L.; Ji, S. H.; Chen, X.; Ma, X.; Zhang, S. C.; Wang, Y.; He, K.; Xue, Q. K. Thickness Dependence of the Quantum Anomalous Hall Effect in Magnetic Topological Insulator Films. *Adv. Mater.* **2016**, *28*, 6386–6390.

(11) Ou, Y.; Liu, C.; Jiang, G.; Feng, Y.; Zhao, D.; Wu, W.; Wang, X.-X.; Li, W.; Song, C.; Wang, L.-L.; Wang, W.; Wu, W.; Zhang, Q.; Gu, L.; Wang, Y.; He, K.; Ma, X.-C.; Xue, Q.-K. Enhancing the Quantum Anomalous Hall Effect by Magnetic Codoping in a Topological Insulator. *Adv Mater* **2017**, *2017*, 1703062.

(12) He, K.; Wang, Y.; Xue, Q. K. Quantum Anomalous Hall Effect. *Natl. Sci. Rev.* **2014**, *1* (1), 38–48.

(13) Zhang, J.; Chang, C.-Z.; Tang, P.; Zhang, Z.; Feng, X.; Li, K.; Wang, L. -l.; Chen, X.; Liu, C.; Duan, W.; He, K.; Xue, Q.-K.; Ma, X.; Wang, Y. Topology-Driven Magnetic Quantum Phase Transition in Topological Insulators. *Science* **2013**, *339* (6127), 1582–1586.

(14) Zhang, L.; Zhao, D.; Zang, Y.; Yuan, Y.; Jiang, G.; Liao, M.; Zhang, D.; He, K.; Ma, X.; Xue, Q. Ferromagnetism in Vanadium-Doped $Bi_2Se_3$ Topological Insulator Films. *APL Mater.* **2017**, *5* (7), 076106.

(15) Kim, J.; Jhi, S.-H.; MacDonald, A. H.; Wu, R. Ordering Mechanism and Quantum Anomalous Hall Effect of Magnetically Doped Topological Insulators. *Phys. Rev. B* **2017**, *96* (14), 140410.

(16) Chang, C. Z.; Tang, P.; Wang, Y. L.; Feng, X.; Li, K.; Zhang, Z.; Wang, Y.; Wang, L. L.; Chen, X.;

Liu, C.; Duan, W.; He, K.; Ma, X. C.; Xue, Q. K. Chemical-Potential-Dependent Gap Opening at the Dirac Surface States of $Bi_2Se_3$ Induced by Aggregated Substitutional Cr Atoms. *Phys. Rev. Lett.* **2014**, *112* (5), 056801.

(17) Koirala, N.; Brahlek, M.; Salehi, M.; Wu, L.; Dai, J.; Waugh, J.; Nummy, T.; Han, M. G.; Moon, J.; Zhu, Y.; Dessau, D.; Wu, W.; Armitage, N. P.; Oh, S. Record Surface State Mobility and Quantum Hall Effect in Topological Insulator Thin Films via Interface Engineering. *Nano Lett.* **2015**, *15* (12), 8245–8249.

(18) Salehi, M.; Shapourian, H.; Koirala, N.; Brahlek, M. J.; Moon, J.; Oh, S. Finite-Size and Composition-Driven Topological Phase Transition in $(Bi_{1-x}In_x)_2Se_3$ Thin Films. *Nano Lett.* **2016**, *16* (9), 5528–5532.

(19) Moon, J.; Koirala, N.; Salehi, M.; Zhang, W.; Wu, W.; Oh, S. Solution to the Hole-Doping Problem and Tunable Quantum Hall Effect in $Bi_2Se_3$ Thin Films. *Nano Lett.* **2018**, *18* (2), 820–826.

(20) West, D.; Sun, Y. Y.; Wang, H.; Bang, J.; Zhang, S. B. Native Defects in Second-Generation Topological Insulators: Effect of Spin-Orbit Interaction on $Bi_2Se_3$. *Phys. Rev. B* **2012**, *86* (12), 121201(R).

(21) Van Vleck, J. H. Models of Exchange Coupling in Ferromagnetic Media. *Rev. Mod. Phys.* **1953**, *25* (1), 220–227.

(22) Ruderman, M. A.; Kittel, C. Indirect Exchange Coupling of Nuclear Magnetic Moments by Conduction Electrons. *Phys. Rev.* **1954**, *96* (1), 99–102.

(23) Liu, Q.; Liu, C. X.; Xu, C.; Qi, X. L.; Zhang, S. C. Magnetic Impurities on the Surface of a Topological Insulator. *Phys. Rev. Lett.* **2009**, *102* (15), 156603.

(24) Chang, C.-Z.; Zhang, J.; Liu, M.; Zhang, Z.; Feng, X.; Li, K.; Wang, L.-L.; Chen, X.; Dai, X.; Fang,

Z.; Qi, X.-L.; Zhang, S.-C.; Wang, Y.; He, K.; Ma, X.-C.; Xue, Q.-K. Thin Films of Magnetically Doped Topological Insulator with Carrier-Independent Long-Range Ferromagnetic Order. *Adv. Mater.* **2013**, *25* (7), 1065–1070.

(25) Li, M.; Chang, C. Z.; Wu, L.; Tao, J.; Zhao, W.; Chan, M. H. W.; Moodera, J. S.; Li, J.; Zhu, Y. Experimental Verification of the van Vleck Nature of Long-Range Ferromagnetic Order in the Vanadium-Doped Three-Dimensional Topological Insulator $Sb_2Te_3$. *Phys. Rev. Lett.* **2015**, *114* (14), 146802.

(26) Kou, X.; Lang, M.; Fan, Y.; Jiang, Y.; Nie, T.; Zhang, J.; Jiang, W.; Wang, Y.; Yao, Y.; He, L.; Wang, K. L. Interplay between Different Magnetisms in Cr-Doped Topological Insulators. *ACS Nano* **2013**, *7* (10), 9205–9212.

(27) Wang, W.; Ou, Y.; Liu, C.; Wang, Y.; He, K.; Xue, Q.-K.; Wu, W. Direct Evidence of Ferromagnetism in a Quantum Anomalous Hall System. *Nat. Phys.* **2018**, *14* (8), 791–795.

# Supporting information

# Ferromagnetic anomalous Hall effect in Cr-doped $Bi_2Se_3$ thin films via surface-state engineering

*Jisoo Moon[†], Jinwoong Kim[†], Nikesh Koirala[†#], Maryam Salehi[‡], David Vanderbilt[†], Seongshik Oh[†]*

[†]Department of Physics and Astronomy, Rutgers, The State University of New Jersey, Piscataway, New Jersey 08854, U.S.A.

[‡]Department of Materials Science and Engineering, Rutgers, The State University of New Jersey, Piscataway, New Jersey 08854, U.S.A.

**Content**

**I. The anomalous Hall effect in a thinner film: 1-6-1 QL structure of $Ca_{0.04}(Cr_{0.5}Bi_{0.5})_{1.96}Se_3$ – $Ca_{0.04}Bi_{1.96}Se_3$ – $Ca_{0.04}(Cr_{0.5}Bi_{0.5})_{1.96}Se_3$ films**

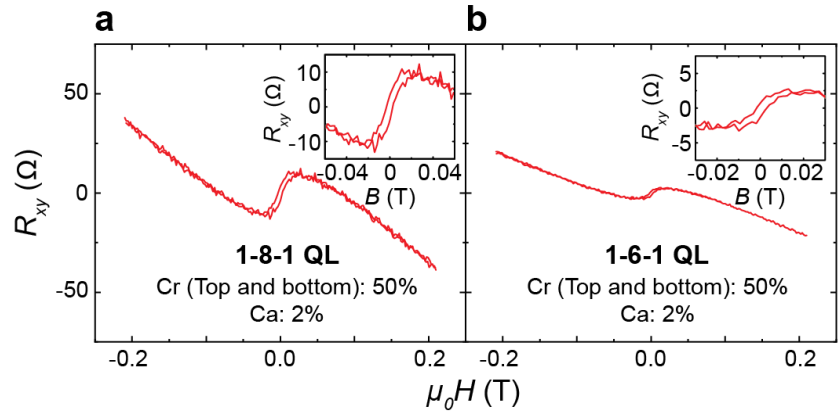

**Figure S1. The anomalous Hall effect signals in the surface engineered films.** Hysteresis loops observed in (a) 1-8-1 QL and (b) 1-6-1 QL structure of $Ca_{0.04}(Cr_{0.5}Bi_{0.5})_{1.96}Se_3$ – $Ca_{0.04}Bi_{1.96}Se_3$ – $Ca_{0.04}(Cr_{0.5}Bi_{0.5})_{1.96}Se_3$ films.

## II. Calculation methods

First-principles calculations are carried out by using VASP[1,2] and wannierized by using the VASP-WANNIER90[3] interface to arrive at a tight-binding description of first-principles quality. The pseudopotential is of the projector-augmented-wave type as implemented in VASP[4,5]. The Perdew-Burke-Ernzerhof approximation for solid (PBEsol)[6] is employed to describe the generalized gradient approximation type of exchange-correlation functional. We choose an energy cutoff of 300 eV and the Brillouin zone is sampled with k-point grids of size $10 \times 10 \times 10$ for pristine $Bi_2Se_3$ and $4 \times 4 \times 1$ for $\sqrt{3} \times \sqrt{3} \times 1$ hexagonal supercell structures in 3 QLs with Cr partially substituting for Bi. Atomic spin-orbit-coupling is added to Bi $p$ and Se $p$ orbital basis after wannier projection.

## III. Details for the mass-gap analysis

In Fig. 4a and b, we find that our computed Dirac cone splittings $E_g$ are well described by the formula

$$E_g = \min | E_{mag} \pm E_0 |, \quad (1)$$

where $E_0 = 19\ \mu eV$ is the gap induced by the interaction between top and bottom surfaces of an 8 QL thick film, and

$$E_{mag} = \chi_{Bi} \Delta_{Bi} + \chi_{Se} \Delta_{Se} \quad (2)$$

with $\chi_{Bi} = 0.420$ and $\chi_{Se} = -0.034$ being the mass-gap susceptibilities defined as the ratios of the induced mass gap to the applied Zeeman field. Two topological transitions then occur when $E_g = 0$, with the film as a whole exhibiting Chern numbers of $C = -1, 0$, and $+1$ for $E_{mag} > E_0$, $|E_{mag}| < E_0$, and $E_{mag} < -E_0$, respectively. The mass-gap susceptibility of Se is, in principle, expressed as $\chi_{Se} = 1/3(\chi_{Se1} + 2\chi_{Se2})$, where Se1 and Se2 refer to the central and the two outer Se atoms, respectively. However, the mass-gap susceptibility of Se1 is negligible ($|\chi_{Se1}| < 0.2\ |\chi_{Se2}|$), so the Zeeman splitting of Se1 is not included in the resulting average.

In Fig. 4a and b, the Chern numbers are obtained from the Hall conductivities calculated by the Kubo formula with the Wannier functions as follows.

$$\sigma_{AH,ij} = e^2 \hbar \sum_{n \neq n'} \int \frac{d\mathbf{k}}{(2\pi)^d} [f(\varepsilon_n(\mathbf{k})) - f(\varepsilon_{n'}(\mathbf{k}))] \times \text{Im} \frac{\langle n, \mathbf{k}|v_i|n', \mathbf{k}\rangle\langle n', \mathbf{k}|v_j|n, \mathbf{k}\rangle}{[\varepsilon_n(\mathbf{k}) - \varepsilon_{n'}(\mathbf{k})]^2} \quad (3)$$

$f$ is the Fermi distribution function, and $\varepsilon_n(\mathbf{k})$ is an energy eigenvalue of the $n^{th}$ band at $\mathbf{k}$. $v_i$ and $v_j$ are the velocity operators of $i$ and $j$ cartesian directions. $d$ is the dimensionality, so $d = 2$ due to the 2D nature of the topological surface states. The Berry curvature is sampled in a two-dimensional $k$-point mesh whose grid density corresponds to $100 \times 100$ near the Brillouin zone boundary and $10^5 \times 10^5$ near the $\Gamma$ point. The Fermi level is set to lie in the direct band gap so that the Chern number would be quantized.

REFERENCE

(1) Kresse, G.; Furthmüller, J. Efficient Iterative Schemes for Ab Initio Total-Energy Calculations Using a Plane-Wave Basis Set. *Phys. Rev. B* **1996**, *54* (16), 11169–11186.

(2) Kresse, G.; Furthmüller, J. Efficiency of Ab-Initio Total Energy Calculations for Metals and Semiconductors Using a Plane-Wave Basis Set. *Comput. Mater. Sci.* **1996**, *6* (1), 15–50.

(3) Mostofi, A. A.; Yates, J. R.; Pizzi, G.; Lee, Y.-S.; Souza, I.; Vanderbilt, D.; Marzari, N. An Updated Version of Wannier90: A Tool for Obtaining Maximally-Localised Wannier Functions. *Comput. Phys. Commun.* **2014**, *185* (8), 2309–2310.

(4) Blöchl, P. E. Projector Augmented-Wave Method. *Phys. Rev. B* **1994**, *50* (24), 17953–17979.

(5) Kresse, G.; Joubert, D. From Ultrasoft Pseudopotentials to the Projector Augmented-Wave Method. *Phys. Rev. B* **1999**, *59* (3), 1758–1775.

(6) Perdew, J. P.; Ruzsinszky, A.; Csonka, G. I.; Vydrov, O. A.; Scuseria, G. E.; Constantin, L. A.; Zhou, X.; Burke, K. Restoring the Density-Gradient Expansion for Exchange in Solids and Surfaces. *Phys. Rev. Lett.* **2008**, *100* (13), 136406.